\documentclass[letterpaper,10pt]{article} 

\usepackage{opticameet3} 

\newcommand\authormark[1]{\textsuperscript{#1}}

\usepackage{amsmath,amssymb}
\usepackage[colorlinks=true,bookmarks=false,citecolor=blue,urlcolor=blue]{hyperref} 
\usepackage[acronym]{glossaries}
\usepackage{relsize}
\usepackage{lipsum}
\usepackage{multicol}
\usepackage{etoolbox}

\newacronym{qkd}{QKD}{quantum-key distribution}
\newacronym{skr}{SKR}{secret key rate}
\newacronym{wdm}{WDM}{wavelength-division multiplexing}
\newacronym{ase}{ASE}{amplified spontaneous emission}
\newacronym{sprs}{SpRS}{spontaneous Raman scattering}
\newacronym{fwm}{FWM}{four-wave-mixing}
\newacronym{psd}{PSD}{power spectral density}
\newacronym{cw}{CW}{continuous wave}
\newacronym{tls}{TLS}{tunable light source}
\newacronym{ws}{WS}{wave shaper}
\newacronym{osa}{OSA}{optical spectrum analyzer}
\newacronym{smf}{SMF}{single-mode fiber}

\begin{document}

\title{Experimental Characterization and Model Validation of Interference in Classical-QKD Coexistence Transmission}
\copyrightyear{2025}

\author{Lucas~Alves~Zischler\authormark{1,*}, Amirhossein~Ghazisaeidi\authormark{2}, Carina~Castiñeiras~Carrero\authormark{2}, Tristan~Vosshenrich\authormark{2},~Jeremie~Renaudier\authormark{2},~Antonio~Mecozzi\authormark{1},~and~Cristian~Antonelli\authormark{1}}

\address{
  \authormark{1}Department of Physical and Chemical Sciences, University of L’Aquila, L’Aquila, Italy\\
  \authormark{2}Optical Transmission Department, Nokia Bell Labs, Massy, France
}

\email{\authormark{*}Corresponding author: lucas.zischler@univaq.it}

\vspace{-0.2cm}
\begin{abstract}
We present an experimental characterization of coexistence-induced interference in QKD transmission arising from SpRS and FWM, validating a comprehensive semi-analytical model for accurate noise estimation. Experimental results show good agreement with theoretical predictions.
\end{abstract}

\vspace{0.2cm}
\section{Introduction}
By exploiting fundamental principles of quantum mechanics, \gls*{qkd} enables physically guaranteed secure key exchange, with growing interest for practical deployment~\cite{lo2014secure}. Most \gls*{qkd} protocols encode information in photons, allowing transmission over existing optical fiber networks. Although maximum \glspl*{skr} are achieved using dedicated dark fibers, quantum and classical signals can coexist within the same optical link to optimize the infrastructure usage.

\Gls*{wdm} transmission offers flexibility in the design of coexistence schemes, specially while expanding to multi-band scenarios. Nevertheless, the significant power disparity between classical and quantum channels can severely degrade \gls*{qkd} performance through non-linear effects and \gls*{ase} noise. While the latter can be suppressed through filtering, non-linear effects are unmitigable in silica-core fibers. The most relevant non-linear impairments are \gls*{sprs} and \gls*{fwm}. The \gls*{sprs} is a broadband bidirectional effect, and band separation between classical and quantum channels is required to mitigate its interference. In contrast, \gls*{fwm} is a narrowband effect, and can become dominant in transmission scenarios where the quantum channel is placed adjacent to classical channels and they propagate in the same direction~\cite{peters2009dense}. To predict \glspl*{skr} in coexistence scenarios and to evaluate the impact of different classical transmission configurations, accurate theoretical models for interference noise evaluation are required.

In this work we characterize \gls*{sprs} and \gls*{fwm} noise in transmission settings where they are most relevant, namely, \gls*{sprs} in multi-band transmission, and \gls*{fwm} interference within the C-band. The experimental results are used to successfully validate the comprehensive semi-analytical model that we recently developed~\cite{zischler2025accurate}.

\section{Models for coexistence impairments}

The impact of noise differs across \gls*{qkd} protocols, but can generally be expressed as a function of noise \gls*{psd} and device parameters. The noise affecting the quantum signal can be categorized into three main sources: noise from quantum devices and shot noise, both independent of classical channels; unfiltered \gls*{ase} and other types of leakage from the classical channels at the transmitter, which can impact the quantum channel even in counter-propagation schemes via Rayleigh backscattering; non-linear interference generated within the fiber, primarily due to \gls*{sprs} and \gls*{fwm} in practical coexistence schemes. Theoretical models for estimating the \gls*{psd} of the noise generated by non-linear effects exist in the literature~\cite{du2020impact, zischler2025accurate}, and rely on the knowledge of the classical power spectrum and fiber parameters, including \gls*{sprs} efficiency and attenuation profiles. The model utilized here is provided in detail in~\cite[Eq.~(10)--(33)]{zischler2025accurate}.

\section{Experimental Results and Model Validation}

We evaluate sample scenarios to illustrate key effects in coexistence transmission and compare against theoretical predictions. The \Gls*{sprs} is analyzed in multi-band configurations, being the dominant non-linear impairment. While \gls*{fwm} is examined in co-propagation scenarios where the quantum channel is allocated in between high-power \gls*{cw} signals.

\begin{figure}[!h]
    \centering
    \vspace{-0.6cm}
    \includegraphics{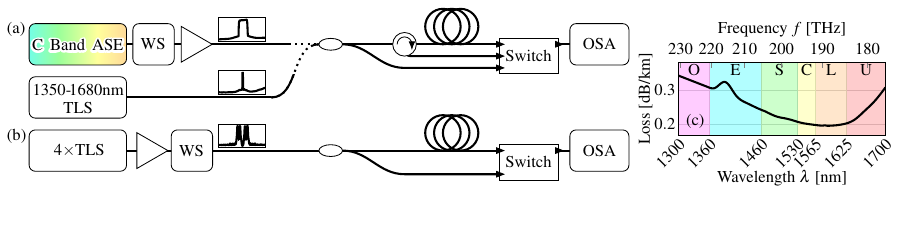}

    \vspace{-1.2cm}
    \caption{Setup diagram for (a) \gls*{sprs} and (b) \gls*{fwm} measurements. (c) Measured attenuation profile.}
    \vspace{-0.8cm}
    \label{fig:Setup}
\end{figure}

The experimental setups for \gls*{sprs} and \gls*{fwm} measurements are shown in Fig.~\ref{fig:Setup}(a) and~\ref{fig:Setup}(b), respectively. To characterize \gls*{sprs} efficiency, a narrow-linewidth \gls*{tls} sweeps the spectrum from 1350~to~1680~nm at 6~dBm launch power. Classical transmission is emulated with a C-band \gls*{ase}-loaded spectrum, flattened through a \gls*{ws} and amplified. The switch allows measurement from co- and counter-propagating noise, and launch power profiles by the \gls*{osa}. Noise from \Gls*{fwm} is measured at 194.7~THz (1539.8~nm) with unmodulated \gls*{cw} tones placed at $\pm$50 and $\pm$100~GHz around the quantum channel, with launched and received profiles measured by the \gls*{osa}. The \gls*{cw} signals are amplified and spectrally filtered by the \gls*{ws}, producing a notch around the quantum channel to suppress crosstalk prior to transmission. All signals propagate through \gls*{smf} spools, with attenuation profile shown in Fig.~\ref{fig:Setup}(c).
\vspace{-0.3cm}

\begin{figure}[!h]
    \centering
    \includegraphics{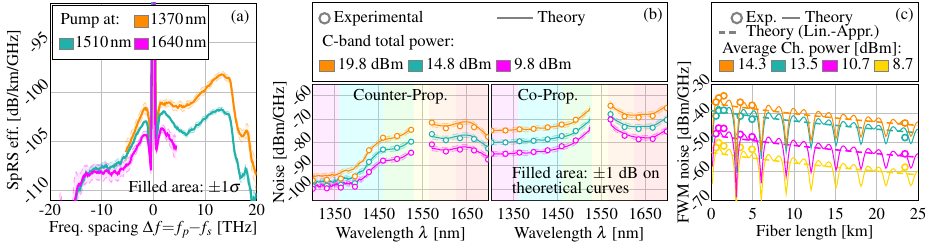}

    \vspace{-0.3cm}
    \caption{(a) \gls*{sprs} efficiency as a function of frequency separation for pump signals at specified wavelengths. (b) Measured multi-band \gls*{sprs} noise. (c) \gls*{fwm} noise versus fiber length.}
    \vspace{-0.5cm}
    \label{fig:Fig}
\end{figure}

Figure~\ref{fig:Fig}(a) shows measured \gls*{sprs} efficiency from backscattered noise. Filled regions indicate $\pm 1$ standard deviation. Consistent with the theory~\cite{lin2007photon}, \gls*{sprs} increases for pumps at lower wavelengths due to the reduced effective area and is higher in the Stokes region. As a consequence of these two factors, \gls*{qkd} schemes where classical signals are allocated in lower frequencies can suffer from \gls*{sprs} impairments even with spectral separations exceeding 200~nm~\cite{beppu2025coexistence}.

Figure~\ref{fig:Fig}(b) shows the measured noise spectra at 50~km, for given full C-band signal power levels, alongside theoretical values estimated with~\cite{zischler2025accurate}. Filled areas represent an uncertainty of $\pm 1$~dB on theoretical estimations. Overall, the theory is in good agreement with measurements, with minor divergences observed in the lower E-band at high C-band loading. In counter-propagation, noise in the O- and half of the E-band plateaus as \gls*{sprs} becomes negligible, reaching the background noise plus Rayleigh backscattered \gls*{ase} (Rayleigh efficiency measured as -42.32~dB/km). In co-propagation, unfiltered \gls*{ase} dominates and notch-filtering is necessary for \gls*{qkd} transmission. In such scenario noise evolves simply according to attenuation values.

Figure~\ref{fig:Fig}(c) shows the measured \gls*{fwm}-induced noise (after subtraction of the linear contribution from measurements) as a function of fiber spool length for given channel powers. Theoretical curves are given for the exact \gls*{fwm} efficiency~\cite[Eq.~(12)]{zischler2025accurate}, and for its linearly-averaged approximation~\cite[Eq.~(22)]{zischler2025accurate}, which enables more efficient numerical computation. Analytical values consider group velocity dispersion of $\beta_{2} = \text{-}21.1~\text{ps}^{\text{2}}/\text{km}$ and non-linearity coefficient of $\gamma = 1.3~\text{W}^{-1}\text{km}^{-1}$. The exact theoretical \gls*{fwm} solution is in great agreement with measurements. Likewise, despite not encompassing the rapid oscillations, the linearly-averaged approximate solution also shows good accuracy throughout the measured dataset.

\section{Conclusion}

We experimentally characterized the primary sources of non-linear interference for \gls*{qkd} coexistence transmission. \gls*{sprs} is the dominant impairment in multi-band coexistence schemes, while \gls*{fwm} can prevail in co-propagation scenarios, where the quantum channel is placed adjacent to classical signals. The experimental results were validated against analytical models, showing strong agreement between both approaches.

\footnotesize
\vspace{0.15cm}
\noindent\textbf{Acknowledgement} This work was supported in part by the European Union's Grant Agreement No. 101120422 - QuNEST.
\vspace{-0.2cm}
\footnotesize
\bibliographystyle{opticajnl.bst}

~\\[-6pt]
\begin{multicols}{2}
~\\[-48pt]

\bibliography{shorter.bib}
\end{multicols}

\end{document}